\documentclass{XrU2005}
\usepackage{graphicx}
\usepackage{natbib}

\title{High spatial resolution mapping of extinction in the SMC with the Swift-UVOT}
\author[1]{A. J. Blustin}
\author[2]{W. B. Landsman}
\author[3]{M. Still}
\author[1]{S. R. Rosen}
\author[1]{M. J. Page}
\author[4]{P. W. A. Roming}
\affil[1]{UCL Mullard Space Science Laboratory, Holmbury St. Mary, Dorking, Surrey, RH5 6NT, UK}
\affil[2]{NASA/GSFC, Code 681, Greenbelt, MD 20771, USA}
\affil[3]{SAAO, PO Box 9, Observatory 7935, South Africa}
\affil[4]{Dept. of Astronomy \& Astrophysics, Pennsylvania State University, 525 Davey Lab, University Park, PA 16802, USA}

\begin{document}


\maketitle

\begin{abstract}
The wide range of UV extinction properties in the Small Magellanic Cloud (SMC) probably traces the effects of star formation on interstellar dust. The Swift UVOT, with its three UV filters (centred at 1800, 2200 and 2500~\AA\ respectively) and three optical filters, is an ideal instrument to map this extinction. We present preliminary results of six-band photometry in a $\sim$ 3.3\arcmin\ square field in the SMC. Fitting the resulting optical/UV spectral distributions with a dust model, we estimate the depths of extinction for 53 stars and use this to map the distribution of dust in the observed field. We briefly discuss the relevance of this study for star forming galaxies, including the hosts of Gamma-Ray Bursts.
\end{abstract}

\section{Introduction and analysis}

It is known that there is a large range of UV extinction properties in the SMC, where the 2200~\AA\ bump and far-UV extinction below 2000~\AA\ varies among the few stars with UV spectroscopic data \citep{gordon2003}. The Swift UVOT \citep{roming2004} is an ideal instrument with which to investigate this, since it has three UV filters sensitive in the bands of interest: UVW2 centred at 1800~\AA, UVM2 at 2200~\AA\ and UVW1 at 2500~\AA\ as well as optical U, B and V filters. We report the results of a pilot study into generating high spatial resolution maps of interstellar extinction in the SMC using Swift UVOT images. We have 17\arcmin\ square images of a Northern pointing in the SMC in all six colour filters. Since these images contain thousands of stars, even in the furthest UV band, for the initial analysis reported here we tested our method in a smaller region. We chose a 200 pixel ($\sim$ 3.3\arcmin) square area in the North-West corner of the field where the background is relatively low in all filters and the image is less crowded. 

The images were aspect-corrected so the stellar positions were consistent both between images and with the Digital Sky Survey. Since the UVOT PSF is $\sim$ 2\arcsec\ FWHM in the ultraviolet, multiple sources in this crowded field could look like a single star. Data from the higher spatial resolution \citet{zaritsky2002} Magellanic Clouds photometric survey were therefore used to check for superposition. With reference to this, 53 isolated stars, distributed over the field and visible in all filters, were selected; very bright sources were excluded to avoid coincidence loss (the UVOT has a photon-counting detector). A single background region was used for each image, at the same position on the sky in each case. The Swift software tool uvot2pha was used to generate Xspec-compatible spectral files and background files for each star in each filter. The errors on the source counts were of order 10\% to take account of systematic uncertainty in the counts to flux conversion.

\begin{figure}[h]
\centering
\includegraphics[width=7cm]{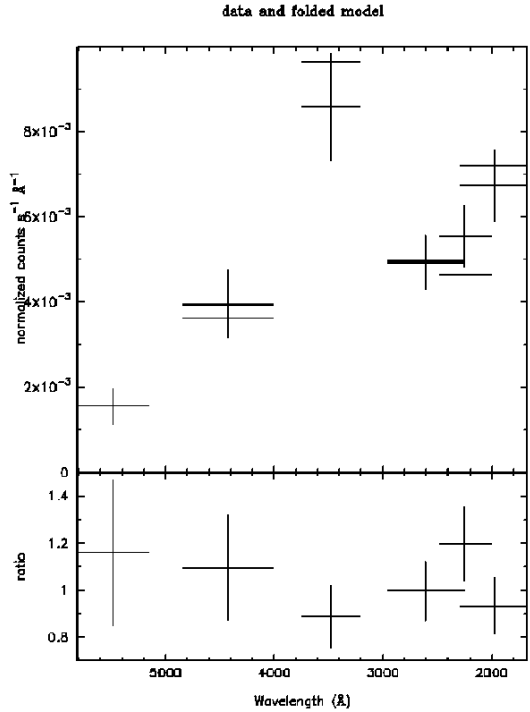}
\caption{UVOT broad-band spectrum of the star with the least extinction in our field, with model overplotted. The lower panel shows the ratio of data to model.}
\label{star_11}
\end{figure}

The resulting optical/UV spectral energy distributions were fitted in Xspec with a model consisting of a blackbody and two layers of dust extinction (see Fig~\ref{star_11}). The blackbody was used to represent the stellar continuum, with a starting temperature of 0.05~keV as appropriate for a hot O star. Two dust components were used; one for the Milky Way and one for the SMC. The dust model allows the user to specify the type of dust (Milky Way, LMC or SMC) and R$_{\rm v}$, the ratio of total to selective extinction. We assume that R$_{\rm v}$ = 2.93 for the SMC and R$_{\rm v}$ = 3.08 for the Milky Way \citep{pei1992}. The E(B-V) of the local component was fixed at 0.05 \citep{bessell1991}, and its redshift at zero. For the SMC dust component, we used a fixed value of 0.00053 for the redshift \citep{richter1987} and allowed E(B-V) to vary. In all but eight cases, the fits were relatively good with a reduced $\chi^{2}$ less than 2. The residuals in most of the fits are of a similar form, showing that there is a systematic error in the modelling which is not just limited to, say, poor photometry of a particular star. More accurate modelling of the stellar continuum is almost certainly required for future work, since the spectra of such hot stars will not be a straightforward blackbody due to the presence of the Balmer discontinuity. 

\section{Results and discussion}

\begin{figure}[h]
\centering
\includegraphics[width=7cm]{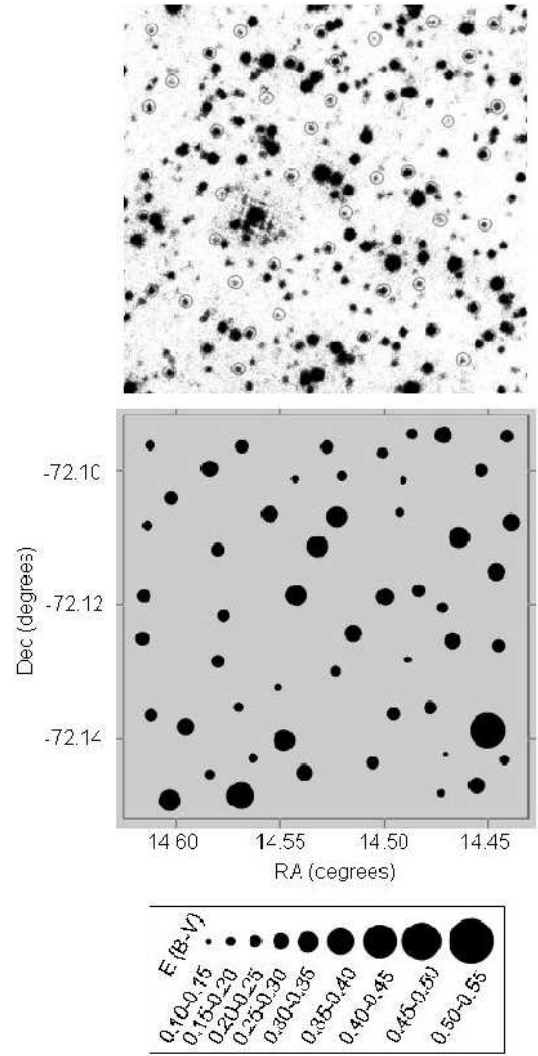}
\caption{(Top) UVW1 (2500~\AA) image of a 3.3\arcmin\ square field in the SMC, with circles marking the stars used in this analysis, and (bottom) intrinsic extinction map of the same field.}
\label{stars_ext_vert}
\end{figure}

Fig.~\ref{stars_ext_vert} shows an image of the $\sim$ 3.3\arcmin\ square field alongside a map of E(B-V). There does indeed appear to be quite a variation in extinction between individual stars, although the elucidation of any underlying patterns or correlations with locations and ages of stars will require a much larger map. Further work will involve making the photometry and background subtraction more sophisticated, obtaining a more appropriate stellar model for the fits, and then constructing an extinction map of the whole of the SMC field observed by the UVOT, preferably with more closely-spaced stars to increase the spatial resolution of the map. It will be possible to extend the mapping to more regions in the SMC and also the LMC. The final maps of dust extinction will be a useful resource to those studying high-energy transient sources in the Magellanic Clouds. Also, variations in UV extinction are related to the effects of star formation on interstellar dust, so comparing our dust maps with the ages and distribution of stars in the field will provide detailed information for studies of star formation in the SMC. Finally, understanding the distribution of dust in a star forming galaxy like the SMC will help in interpreting the intrinsic extinction properties of GRBs, perhaps providing an indication of the stellar environment of the GRB precursor; GRBs do not show the 2200~\AA\ dip that we see in Galactic dust, so we expect them to have dust like that in the SMC.

\end{document}